\documentclass[a4paper,12pt]{article}

\usepackage{graphicx}

\def\pzp{p_0^{\prime}}
\def\pmp{p^{\prime}}

\def\pp{{\bar p}}

\def\im{{\rm Im\,}}
\def\re{{\rm Re\,}}

\def\lc{L}
\def\lcp{{L}^{\prime}}
\def\lcpp{{L}^{\prime\prime}}

\begin{document}

\begin{center}
{\large\bf The rank-one separable interaction kernel for nucleon with scalar propagators}

\vskip 5mm

{S. G. Bondarenko$^{1,\P}$}, {V. V. Burov$^{1}$} , {S. A. Yurev$^{1}$}

\vskip 5mm

$^1${Joint Institute for Nuclear Research, Dubna, 141980, Russia}

\vskip 5mm

$^\P${E-mail: bondarenko@jinr.ru}
\end{center}

\centerline{\bf Abstract}
In the paper the covariant kernel of the nucleon-nucleon
interaction of particles with scalar propagators is analyzed.
The Bethe-Salpeter equation for the $T$ matrix is considered
in the rank-one separable kernel.
The parameters of the kernel for the specific partial-wave
channels explicitly connect with the observables --
low energy scattering parameters and phase
shifts, deuteron binding energy.
Covariant separable kernels forthe  partial-wave channels
with total angular momentum
$J=0$ ($^1S_0$, $^3P_0$) and $J=1$ ($^3S_1- ^3D_1$,
$^1P_1$, $^3P_1$) are constructed.\\
PACS: 11.10.St, 13.75.Cs.

\section{Introduction}
Investigation of few-nucleon systems is very important
for understanding the strong interactions.
For a consistent covariant description of the nucleon-nucleon ($NN$)
interactions the relativistic Bethe-Salpeter (BS) equation~\cite{BSE}
is commonly used.
The BS approach with the separable kernel of interaction
gives a good description of elastic and inelastic
electromagnetic processes with the deuteron~\cite{report}.
For instance, the BS formalism facilitates an analysis of the role of P-
waves (negative energy partial-wave components of the BS amplitude)
in the electromagnetic properties of the deuteron and its comparison with the
nonrelativistic treatment~\cite{Pstates-1999}. Furthermore, the covariant BS approach makes
it possible to analyze off-mass-shell effects and contributions
of the relativistic two-body currents~\cite{tbc-2018}.

Another important topic of the physics of strong interaction is studding
the three-nucleon systems and hadron-deuteron reactions.
The relativistic three-particle systems are described by the Faddeev equations within
the BS approach -- the so called Bethe-Salpeter-Faddeev equations.
A general consideration of the three-nucleon systems is a rather
difficult task. So some simplification should be done at the first
stage of such investigations.

First of all the only pair-interactions are taken into account while
the three-body forces are considered.
Second is using the covariant separable kernel of $NN$
interaction. In this case the system of the 6-fold integral equations
transfer to system of the 2-fold integral equations with generalized
potentials. The third assumption is to treat 
all nucleons having equal masses and the scalar propagators instead of spinor ones.
The spin-isospin structure of the nucleons
is taken into account by using the so-called recoupling-coefficient matrix.

In the paper the covariant kernel of the $NN$
interaction of the particles with scalar propagators for the
partial-wave channels with the total angular momentum $J=0,1$
is analyzed. Obtained results will be used in the future calculations
of the three-nucleon systems.

The paper is organized as following: in sec. 2 the formalism is given,
the details of the calculations and results are presented in sec. 3 and
conclusion is given in sec. 4.

\section{Formalism}
The Bethe-Salpeter equation
for the nucleon-nucleon $T$ matrix is written as
\begin{eqnarray}
T({\hat p}^\prime, {\hat p}; {\hat P}) =
V({\hat p}^\prime, {\hat p}; {\hat P})
+ \frac{i}{4\pi^3} \int\, d^4{\hat k}\,
V({\hat p}^\prime, {\hat p}; P) S({\hat k};{\hat P}) T({\hat k}, {\hat p}; {\hat P}).
\label{t01a}\end{eqnarray}
Here the total four-momentum ${\hat P}=({\hat p}_1+{\hat p}_2)$ and the relative
four-momentum ${\hat p}=({\hat p}_1-{\hat p}_2)/2$ [${\hat p}^{\prime}=({\hat p}^{\prime}_1-{\hat p}^{\prime}_2)/2$]
are introduced,
and $V$ is the kernel of $NN$ interaction (for details, see reference~\cite{report}).

The nucleons in the equation are treated as spin-one-half particles with
scalar propagators
\begin{eqnarray}
S({\hat k};{\hat P}) = [({\hat P}/2+{\hat k})-m_N^2+i0]^{-1}[({\hat P}/2-{\hat k})-m_N^2+i0]^{-1}.
\label{s01}\end{eqnarray}

The partial-wave decomposition of the equation
(in the rest frame of the two-nucleon system)
leads to the following form: 
\begin{eqnarray}
t_{\lcp\lc}(\pzp, \pmp, p_0, p; s) &=&
v_{\lcp\lc}(\pzp, \pmp, p_0, p; s)
\label{t01}\\
&+& \frac{i}{4\pi^3} \sum_{\lcpp} \int\, dk_0\,\int\,
k^2\, dk\,
v_{\lcp\lcpp}(\pzp, \pmp, k_0, k; s)S(k_0,k;s)t_{\lcpp\lc}(k_0, k, p_0, p; s).
\nonumber\end{eqnarray}
Here $s=P^2$, $t$ and $v$ are the partial-wave decomposed $T$ matrix and
kernel $V$ and $e_k=\sqrt{k^2+m^2}$.
For the singlet (uncoupled triplet) case ($L=J$) there is only one term in the sum
and there are two terms for the coupled triplet case ($L=J\mp1$).

To solve the equation~(\ref{t01}) the separable form (rank-one) for the partial-wave decomposed
kernels of interactions is assumed
\begin{eqnarray}
v_{\lcp\lc}(\pzp, \pmp, p_0, p; s) = \lambda g^{[\lcp]}(\pzp, \pmp) g^{[\lc]}(p_0, p)
\label{t04}\end{eqnarray}
where $\lambda$ and $g$ are the parameters and form factors of the model.
Then the $T$ matrix can be written as
\begin{eqnarray}
t_{\lcp\lc}(\pzp, \pmp, p_0, p; s) = \tau(s) g^{[\lcp]}(\pzp, \pmp) g^{[\lc]}(p_0, p),
\label{t05}\end{eqnarray}
with the function $\tau(s)$ being
\begin{eqnarray}
\tau(s) = 1/(\lambda^{-1} + h(s)),
\label{t06}\end{eqnarray}
while the function $h(s)$ has the following form:
\begin{eqnarray}
h(s) = \sum_{\lc} h_{\lc}(s)= -\frac{i}{4\pi^3}\, \int\, dk_0\,\int\, k^2\, dk\,
\sum_{\lc} [g^{[\lc]}(k_0,k)]^2 S(k_0,k;s).
\label{t07}\end{eqnarray}

To construct the rank-one $NN$ kernel the  
covariant generalization of the {\em Yamaguchi}~\cite{yam}
functions for $g^{[\lc]}(k_0,k)$ are used in the following form
\begin{eqnarray}
g^{[S]}(k_0,k) = \frac{1}{k_0^2-k^2-\beta_0^2+i0},
\label{t11a}
\end{eqnarray}
\vskip -7mm
\begin{eqnarray}
g^{[P]}(k_0,k) = \frac{\sqrt{|-k_0^2+k^2|}}{(k_0^2-k^2-\beta_1^2+i0)^2},
\end{eqnarray}
\vskip -7mm
\begin{eqnarray}
g^{[D]}(k_0,k) = \frac{C_2(k_0^2-k^2)}{(k_0^2-k^2-\beta_2^2+i0)^2},
\label{t11}\end{eqnarray}
where $\beta_L$ and $C_L$ are the model parameters.

\subsection{Deuteron and $NN$-scattering observables}
The on-mass-shell $t(s)$ matrix can be expressed through the following observables:
\begin{enumerate}
  \item
in the singlet (uncoupled triplet) channel
\begin{eqnarray}
t(s) \equiv t(0,\pp,0,\pp,s) = - \frac{8 \pi \sqrt{s}}{\bar p}\,
e^{i\delta}\, \sin{\delta}.
\label{t03}\end{eqnarray}
\item
in the coupled triplet channel
\begin{eqnarray}
t(s) =
 \frac{4 \pi i \sqrt{s}}{\bar p}
\left(
\begin{array}{cc}
\cos{2\epsilon}\ e^{2i\delta_{<}} - 1 &
i\sin{2\epsilon}\ e^{i(\delta_{<}+\delta_{>})} \\
i\sin{2\epsilon}\ e^{i(\delta_{<}+\delta_{>})} &
\cos{2\epsilon}\ e^{2i\delta_{>}} - 1\\
\end{array}
\right),
\label{t10}\end{eqnarray}
\end{enumerate}
with $\pp = \sqrt{s/4-m^2} = \sqrt{mT_{lab}/2}$. We introduced scattering phase shifts
$\delta\equiv\delta_{L=J}$, $\delta_{<}\equiv\delta_{L=J - 1}$, $\delta_{>}\equiv\delta_{L=J + 1}$ and 
mixing parameter $\epsilon$.

If the bound state exists in the partial-wave channel there is a
simple pole on the total momentum squared in the $T$ matrix.
Using eq.~(\ref{t06}) one can write
($M_b=2m_N-E_b$, where $E_b$ is the energy of the bound state):
\begin{eqnarray}
\lambda^{-1} = - h(s=M_b^2).
\label{t10a}\end{eqnarray}

The normalization condition for the deuteron vertex functions can be written
as
\begin{eqnarray}
  \frac{d}{dP_\mu} (h_S(s) + h_D(s)) = 2P_\mu(p_S+p_D)
\label{norm}\end{eqnarray}
where $s=P^2=M_d^2$ and $p_L$ is a partial-wave state pseudoprobabity.

The low-energy parameters -- scattering length $a_{\lc}$ and
effective range $r_{\lc}$ -- are defined
by the following equation
\begin{eqnarray}
\pp^{2\lc+1}\, {\cot}\, \delta_{\lc}(s) =
- 1/a_{\lc} + \frac{r_{\lc}}{2}\pp^2 + {\cal O}(\pp^3).
\label{t03a}\end{eqnarray}

To summarize, the eq.~(\ref{t05}) defines the 
$t$ matrix on the mass-shell ($p_0=p_0^{\prime}=0,p=p^{\prime}=\pp$)
which is related to $NN$-scattering observables and the deuteron --
scattering phase shifts and low-energy parameters, the bound state
energy.

\section{Calculations and results}

The parameters of the model for definite partial-wave channel
-- $\lambda_L$, $\beta_L (C_L)$ -- can be obtained from the
analysis of the $NN$-scattering observables. These values are
calculated from the on-mass-shell $t$ matrix, eq.~(\ref{t05}).
The two-fold integrations in eq.~({\ref{t07}})
can be performed by several ways. In the paper the integration
over the $k_0$ variable is done by using the Cauchy theorem and
the remaining one-fold integration over the $k$ (or $e_k$) variable is done numerically.
As it is shown in~\cite{EPJ-2017} the integration on $k$ 
can be performed only for bound state $\sqrt{s} < 2m_N$ and for
elastic $NN$-scattering with $2m_N < \sqrt{s} < 2(m_N+\beta)$.
So, the considered kinetic energy is restricted to the $T_{lab} < 4\beta$.
      
Considering the $e_k$-integration one find the pole in the point
${\bar e_k} = \sqrt{s}/2$ in function $1/(\sqrt{s}/2-e_k+i0)$
which should be calculated by using the
following formal equation:
\begin{eqnarray}
  \frac{1}{\sqrt{s}/2-e_k+i0}  =
  \frac{P}{\sqrt{s}/2-e_k} -i\pi\delta(\sqrt{s}/2-e_k),
\label{ta12}
\end{eqnarray}  
where the first right hand term gives the integral principal value which
gives the real part of the function $h(s)$.
The second right hand term gives the imaginary part of $h(s)$ in the following form
\begin{eqnarray}
  \im h(s) = \frac{{\bar k}}{16{\bar e_k}\pi}  \sum_L g_L(0,{\bar k})^2,
\label{ta13}
\end{eqnarray}
with ${\bar k} = \sqrt{s/4-m_n^2}$.

The initial values of the parameters for three uncoupled
$P$-states ($^3P_0$, $^1P_1$, $^3P_1$)
and coupled $^3S_1-^3D_1$ states
are taken from our previous paper for the rank-one separable $NN$
interaction kernel for nucleon with spinor propagators~\cite{NPA} and
they are refitted using the scalar nucleon propagators.

The experimental data for the phase shifts are taken from SAID program
http://gwdac.phys.gwu.edu/
and the deuteron energy and low-energy parameter values are taken from ref.~\cite{data}.

\subsection{$^1S_0$ partial-wave state}
The parameters of the $^1S_0$ partial-wave are taken from the
paper~\cite{Tjon-1988} and given in table~{\ref{tab2}} without any changes.
The calculated low-energy properties ($a_s$ and $r_s$) are given
in table~{\ref{tab1}} and the phase shifts is shown in figure~1.

\subsection{$^3P_0$, $^1P_1$, $^3P_1$ uncoupled partial-wave states}
The phase shifts for the uncoupled partial-wave states can be obtained
using the eqs.~(\ref{t05}) and (\ref{t07}), namely
\begin{eqnarray}
  \tan{\delta_L(s)} = \frac{\im{t(s)}}{\re{t(s)}} = \frac{\im{h(s)}}{\lambda_L^{-1}+\re{h(s)}}.
\label{ta14}
\end{eqnarray}
To find the parameters $\lambda_L$ and $\beta_L$ the procedure to minimize the function
\begin{eqnarray}
\chi^2(\lambda_L,\beta_L) = \sum\limits_{i=1}^{n}
\frac{(\delta^{\rm exp}(s_i)-\delta(s_i))^2}
{(\Delta\delta^{\rm exp}(s_i))^2}
\end{eqnarray}
is used where $n$ is the number of the experimental points taken into account.
The initial values of the parameters are taken from the paper~\cite{NPA}.
In calculation of the function $\chi^2$ minimum the kinetic energy maximum is
taken: $T^{max}_{lab}$ = 100 MeV for $^3P_0$-state and
$T^{max}_{lab}$ = 200 MeV for $^1P_1$, $^3P_1$-states.
The calculated parameters are given in table~\ref{tab2} and phase shifts --
in figures~2-4.

\subsection{$^3S_1-^3D_1$ coupled partial-wave states}
The parameters of the $^3S_1-^3D_1$ coupled partial-wave states
are obtained from the parameters in the ref.~\cite{NPA} using
the following procedure:
\begin{itemize}
\item the $C_2$ parameter is refited by using the normalization
  condition (\ref{norm}) to have the $^3D_1$ partial-wave
  state pseudoprobabity $p_D=4,5,6$ \% while the parameters $\beta_0$ and $\beta_2$ are
  fixed
\item the $\lambda$ parameter is refited by using the eq.~(\ref{t10a})
  to have the deuteron binding energy $E_d = 2.2246$ MeV
  while all other parameters are fixed.
\end{itemize}
The calculated parameters and low-energy properties ($a_t$ and $r_t$)
are given in table~\ref{tab3} and phase shifts are presented in figure~5.

\begin{table}[ht]
  \caption{Parameters for the $^1S_0$ partial-wave channel}
  \label{tab1}
\begin{center}
\begin{tabular}{lccc}
	\hline
	& Exp. from~\cite{data} & $^1S_0$ \\
	\hline
$\lambda$ (GeV$^4$) &&-1.12087 \\
$\beta_0$ (GeV)     &&0.228302\\
        \hline
$a_L$ (fm)          &-23.748&-23.753\\
$r_L$ (fm)          &2.75   &2.75\\
	\hline
\end{tabular}
\end{center}
\end{table}

\begin{table}[h]
  \caption{Parameters for the $^{3}P_0$, $^{1}P_1$ and $^{3}P_1$ partial-wave channels}
  \label{tab2}
\begin{center}
\begin{tabular}{lcccccc}
	\hline
	&$^3P_0$ & $^3P_1$ & $^1P_1$ \\
	\hline
	$\lambda$ (GeV$^6$) & 0.0428572& -5.83051 & -3.68029  \\
	$\beta_1$ (GeV) &  0.19904 & 0.48273 & 0.44127 \\
	\hline
\end{tabular}
\end{center}
\end{table}

\begin{table}[h]
  \caption{Parameters for the $^{3}S_1-^{3}D_1$ partial-wave channels}
  \label{tab3}
\begin{center}
\begin{tabular}{lcccccc}
	\hline
	& Exp. from~\cite{data} & $^3S_1-^3D_1$ & $^3S_1-^3D_1$ & $^3S_1-^3D_1$ \\
	&& ($p_d = 4\%$) & ($p_d = 5\%$) & ($p_d = 6\%$)\\
	\hline
$\lambda$ (GeV$^4$) &&-1.83756 & -1.57495 & -1.34207\\
$\beta_0$ (GeV)     &&0.251248 & 0.246713 & 0.242291\\
$C_2$               && 1.71475 & 2.52745 & 3.46353   \\
$\beta_2$ (GeV)     && 0.294096 & 0.324494 & 0.350217\\
        \hline
$a_L$ (fm)          &5.424& 5.454 & 5.454 & 5.453\\
$r_L$ (fm)          &1.756& 1.81  & 1.81  & 1.80\\
	\hline
\end{tabular}
\end{center}
\end{table}
\begin{center}
\begin{minipage}{0.7\textwidth}
  \includegraphics[width=\textwidth]{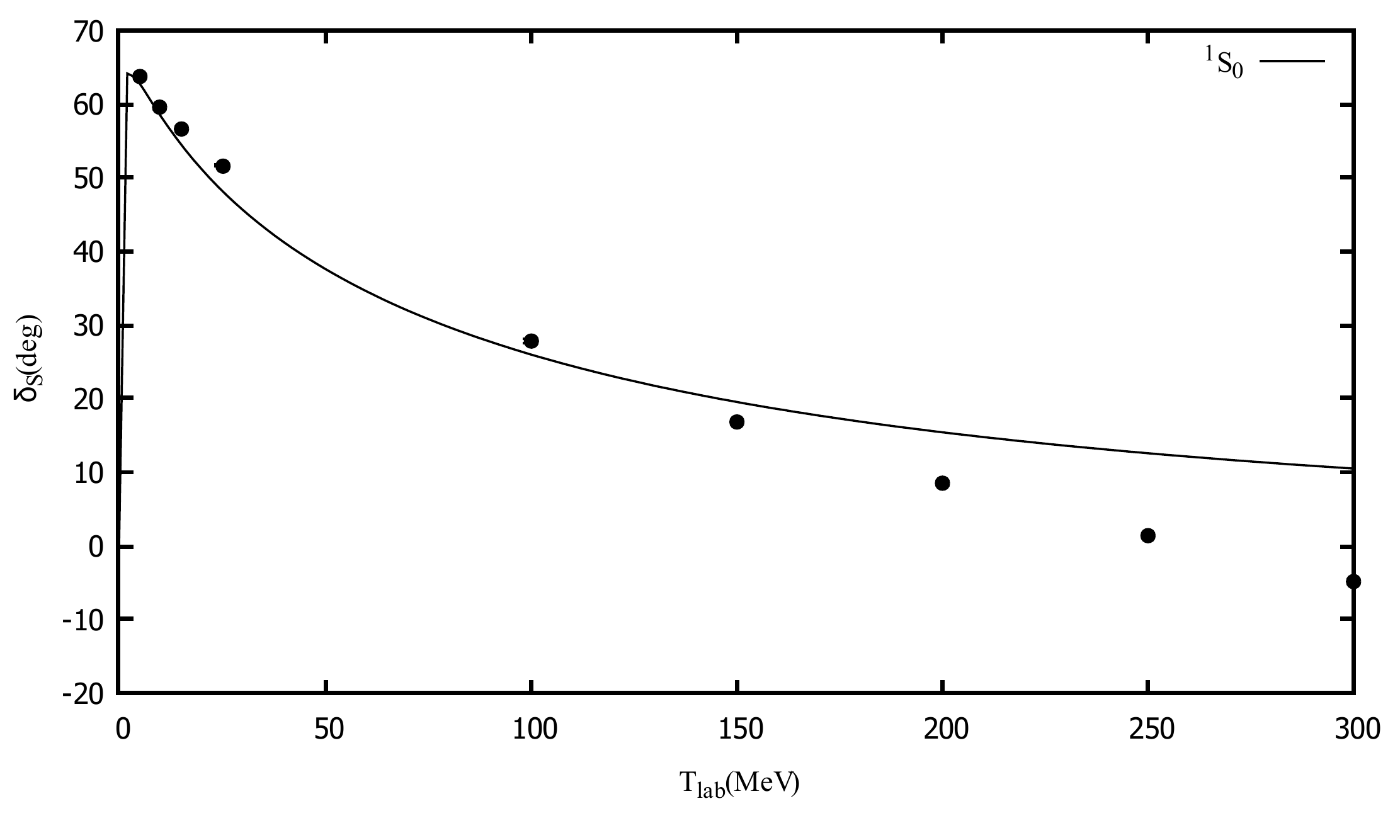}\\[-1mm]
          {Figure 1. The $^1S_0$ channel phase shifts.}
  \label{fig1}
\end{minipage}

\begin{minipage}{0.7\textwidth}
  \includegraphics[width=\textwidth]{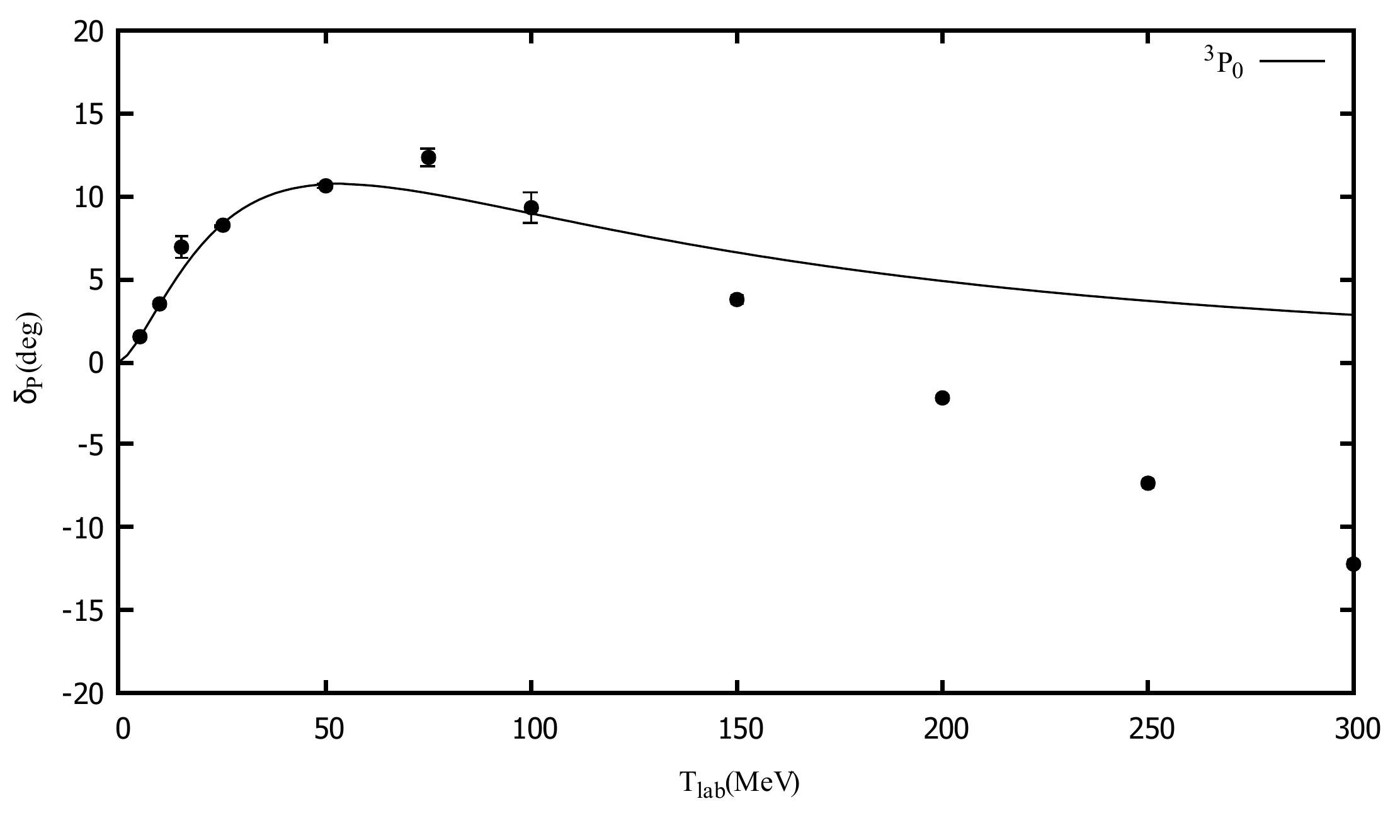}\\[-1mm]  
          {Figure 2. The $^3P_0$ channel phase shifts.}
  \label{fig2}
\end{minipage}

\begin{minipage}{0.7\textwidth}
  \includegraphics[width=\textwidth]{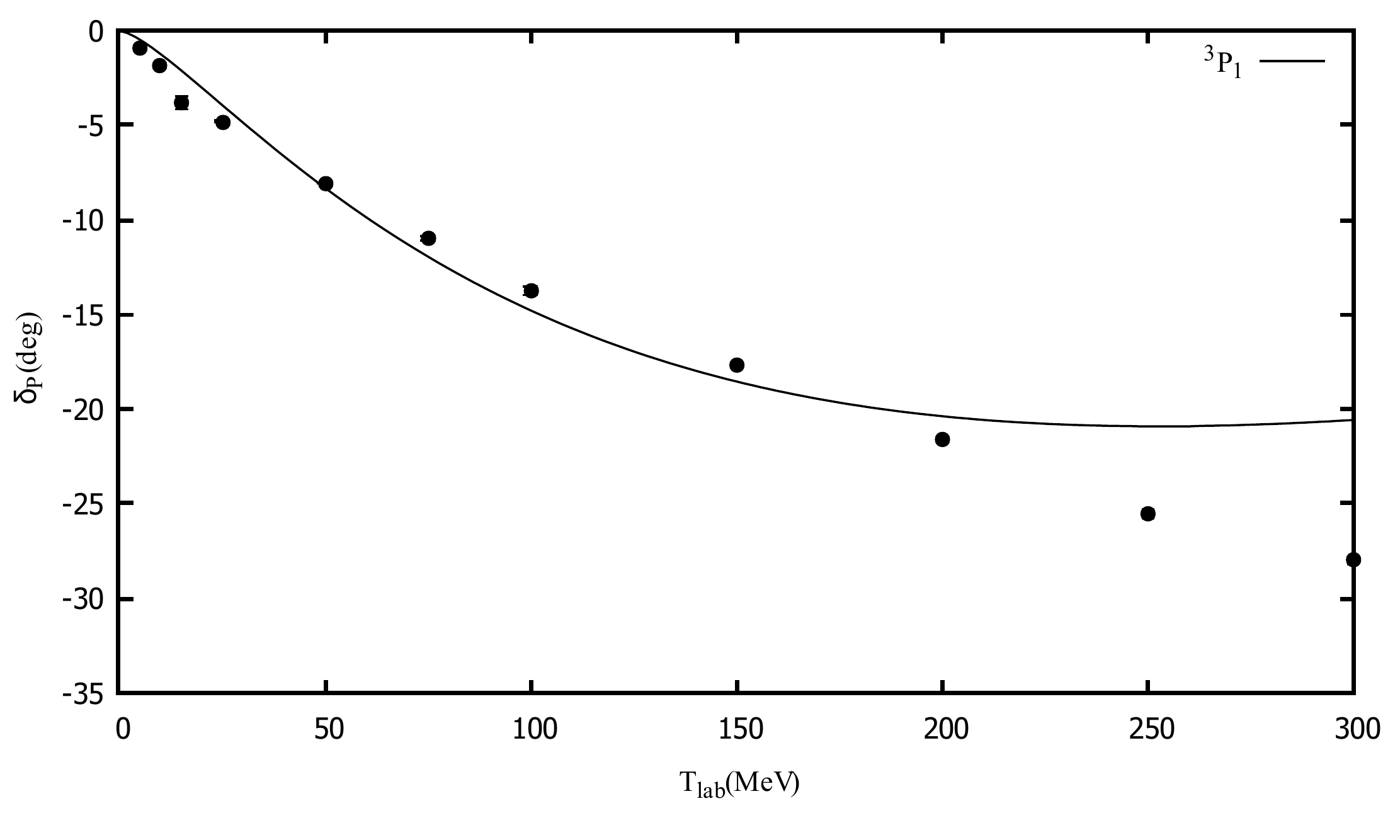}\\[-1mm]  
          {Figure 3. The $^3P_1$ channel phase shifts.}
  \label{fig3}
\end{minipage}

\begin{minipage}{0.7\textwidth}
  \includegraphics[width=\textwidth]{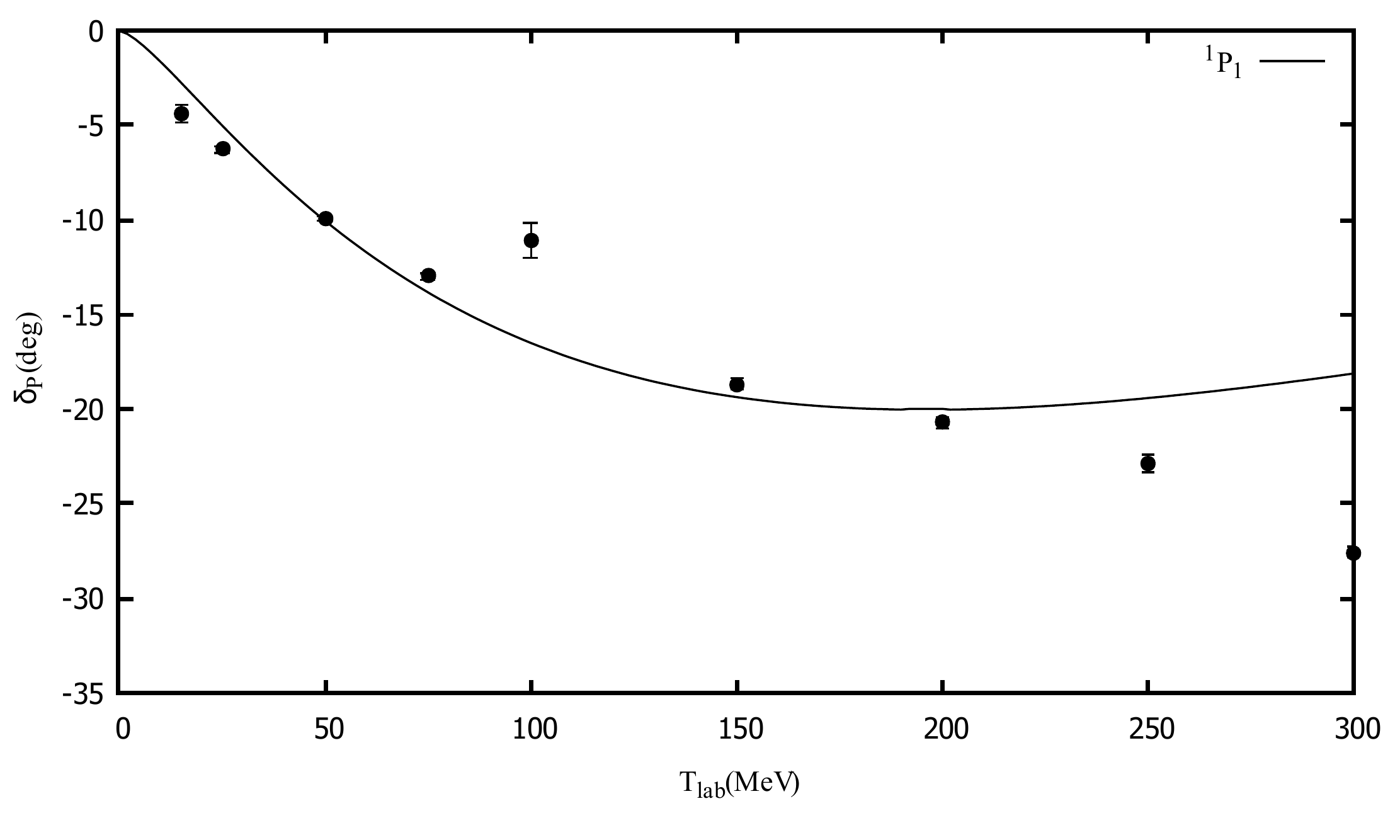}\\[-1mm]  
          {Figure 4. The $^1P_1$ channel phase shifts.}
  \label{fig4}
\end{minipage}

\begin{minipage}{0.7\textwidth}
  \includegraphics[width=\textwidth]{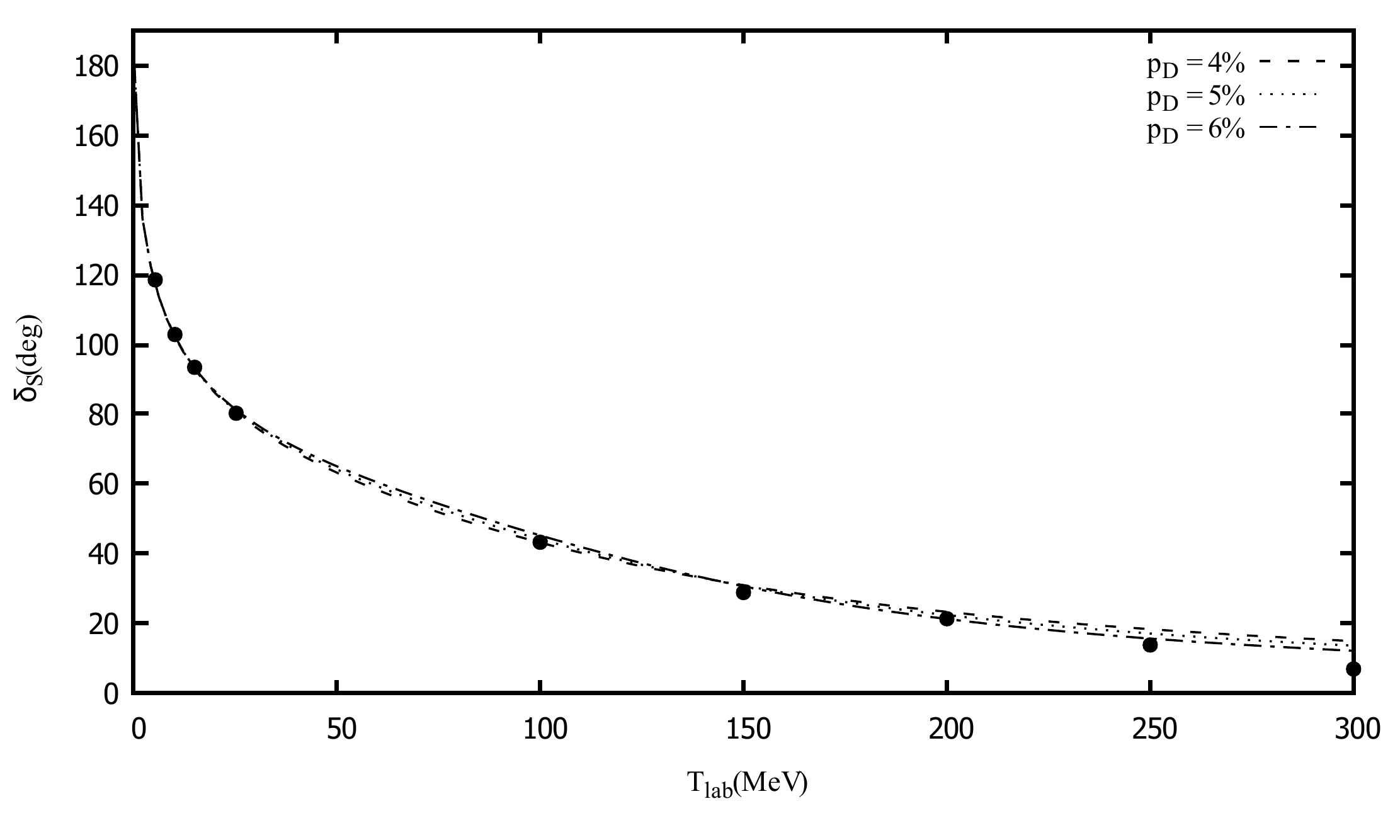}\\[-1mm]
          {Figure 5. The $^3S_1$ channel phase shifts with $p_d=4,5,6$\%.}
  \label{fig5}
\end{minipage}
\end{center}

\section{Conclusion}
The rank-one covariant separable kernels of the $NN$ interaction with
the spin-one-half particles with the scalar propagators are considered.
The parameters for the partial-wave states with the total angular momentum $J=0,1$
are obtained from the analysis of the $NN$-scattering observables
and deuteron static properties.
The proposed models for the kernels will be used to investigate the three-nucleon
bound states and their dynamic electromagnetic properties (form factors).

\section*{Acknowledgment}
This work was partially supported by the Russian Foundation for Basic Research grant No 16-02-
00898.

\end{document}